\documentclass[aip,jap,reprint,graphicx]{revtex4-1} 
\usepackage{graphicx,epsf,epsfig}
\usepackage{dcolumn}
\usepackage{bm}
\usepackage{amsmath, amsthm, amssymb}
\usepackage{color}
\usepackage{ulem}

\usepackage{epstopdf}
\usepackage{float}
\usepackage{longtable}
\usepackage{multirow}

\begin{document}

\title{Instability Mechanism for STT-MRAM switching}
\author{P. B. Visscher}
\affiliation{Center for Materials for Information Technology, U. of Alabama, Tuscaloosa, AL 35401, USA}
\affiliation{Department of Physics and Astronomy, Univ. of Alabama, Tuscaloosa, AL 35401, USA}
\author{Kamaram Munira}
\affiliation{Center for Materials for Information Technology, U. of Alabama, Tuscaloosa, AL 35401, USA}
\author{Robert J. Rosati}
\affiliation{Department of Physics and Astronomy, Univ. of Alabama, Tuscaloosa, AL 35401, USA}
\affiliation{Department of Physics, University of Texas, Austin, TX 78705}
\begin{abstract}
To optimize the design of STT-MRAM (spin-transfer torque magnetic random access memory), it is necessary to be able to predict switching (error) rates.  For small elements, this can be done using a single-macrospin theory since the element will switch quasi-uniformly.  Experimental results on switching rates suggest that elements large enough to be thermally stable switch by some mechanism with a lower energy barrier.  It has been suggested that this mechanism is local nucleation, but we have also previously reported a global magnetostatic instability, which is consistent with the lower experimental energy barriers.  In this paper, we try to determine which of these mechanisms is most important by visualizing the switching in a "U-NU" (uniform - nonuniform) phase diagram.  We find that switching trajectories follow the horizontal U axis (i.e., quasi-uniform precession) until they reach a critical amplitude, at which the magnetostatic instability grows exponentially and a domain wall forms at the center, whose motion completes the switching.  We have tried unsuccessfully to induce local nucleation (a domain wall at the edge).  We conclude that the dominant switching mechanism is not edge nucleation, but the magnetostatic instability.
\end{abstract}

\maketitle

\section{Introduction}
There has been much recent interest in STT-MRAM, in which a magnetic element switches in response to a tunneling current $J$ from a nearby fixed magnetic polarizer.  To design a useful STT-MRAM\cite{MRAM} it is necessary to be able to predict its switching rate.  In particular, the read error rate (the switching rate when $J<J_c$, the critical value for switching) must be small, while the write time must be short (the rate for  $J>J_c$ must be large).  Most previous modeling has been based on the single-macrospin model\cite{chap},\cite{ButlerEtal} in which all of the magnetization vectors are held nearly parallel by the exchange interaction.  This is true when the volume $V$ of the element is small, but then the stability parameter (energy barrier/$k_B T$, or $KV/k_B T$, where $K$ is the anisotropy energy density) is too small for stability (less than about 50).  For elements large enough to be stable, incoherent switching is possible, and experimentally\cite{lowbar} the energy barriers are much less than $KV$, indicating that the switching mechanism is in fact not coherent.  It has been suggested that the mechanism involves local (probably edge) nucleation\cite{actvol}.  However, in previous work\cite{munira}, we have identified a new mechanism for switching, involving a magnetostatic instability. This may be related to the instability of the coherent-precession mode in an ellipsoidal element recently discussed analytically by Bonin \textit{et al}\cite{Bonin}.

In this paper, we use micromagnetic simulation to try to determine which mechanism is most important.  We find that in thermal switching ($J<J_c$) the instability mechanism is dominant.  As we increase the current, nearly-coherent switching becomes more probable, and it dominates in the overdriven case ($J>>J_c$).  We find no regime in which local nucleation is important.

\section{Micromagnetic Model}
We assume a cylindrical STT-MRAM element of thickness $t$ and radius $R$, with perpendicular anisotropy, adjacent to a pinned polarizing layer such that the Landau-Lifshitz (LL) equation for the torque $d\mathbf{M}/dt$ has a "spin torque" proportional to the current:
\begin{equation}\label{LL}
\frac{d\mathbf{M}}{dt} = -\gamma \mathbf{M} \times \mathbf{H} - \frac{\gamma \alpha}{M_s}  \mathbf{M} \times  \mathbf{M} \times \mathbf{H}
- \frac{\gamma J}{M_s}  \mathbf{M} \times  \mathbf{M} \times \mathbf{\hat{m}_p}
\end{equation}
Here $ \mathbf{M}$ is the local magnetization; $ \mathbf{H}$ is the total field, including the exchange, anisotropy, and magnetostatic fields; $M_s$, $\gamma$, $\alpha$ are the saturation magnetization, gyromagnetic factor, and LL damping.  The coefficient $J$ of the spin torque is proportional to current, and has units of magnetic field (kA/m).
The anisotropy field is just $H_K M_z/M_s$, normal to the plane, where $H_K \equiv 2K/ \mu_0 M_s$.  The simulations in this paper were done with our public-domain micromagnetic finite-difference simulator\cite{alamag} -- the magnetizations are defined on a cubic lattice, the exchange field is a linear combination of neighboring magnetizations, and the magnetostatic field is computed using the Fast Multipole Method (FMM)\cite{FMM}. We have omitted terms in $\alpha ^2$ since $\alpha$ is small.

\section{U-NU (Uniform-Nonuniform) phase space}
\label{section:U-NU}
In the single-macrospin picture, the progress of switching is normally described by the angle $\theta$ between the magnetic moment and the film normal.  In a multi-macrospin system, $\theta$ is not uniform, but the normalized transverse part of the squared total magnetic moment
\begin{equation}\label{U}
U = \frac{m_\bot^2}{m_s^2}  \textrm{,    where   } m_\bot^2 \equiv m_x^2+m_y^2
\end{equation}
can be used instead (in the uniform case, $U = \sin^2 \theta$).  If the motion is fairly coherent, $U$ is all we need to describe it.  In terms of the normal modes of the system, $U$ is the (squared) amplitude of the lowest normal mode.  To describe the system completely, we would need the amplitudes and phases of all the normal modes, but we expect the lowest-frequency normal modes will be most important.  

\begin{figure}[!htb]
\begin{center}
\vspace{0.3 in} 
\includegraphics[width=3 in]{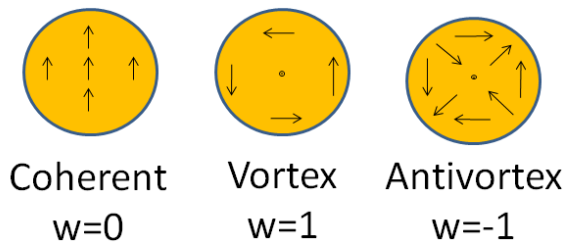}
\end{center}
\vspace{-0.2 in}
\caption{\label{figure:Modes}
The lowest three normal modes of the cylinder, labeled by winding number $w$.}
\end{figure}

In a previous paper\cite{munira}, we classified these modes according to their angular, radial, and $z$ dependence; in the present section, we will briefly summarize these results and indicate how they allow us to quantify the uniform and non-uniform motions.  The angular dependence is most easily described using the complex notation in which a vector $\bf{F}$ in the $xy$ plane is represented by the complex number $\tilde F = F_x  + iF_y$.  The angular dependence of the modes is conveniently labeled by a winding number $w$, the number of times the vector $\bf{F}$ rotates when the point $(x,y)$ rotates once about the $z$ axis.  The simplest function with winding number $w$ is
\begin{eqnarray}\label{Fw}
\tilde F_w (x,y,z) & =   \left( {x + iy} \right)^w  \hspace{0.4 in} & (w>=0) \nonumber \\ 
\tilde F_w (x,y,z) & \hspace{-0.3 in} =   \left( {x - iy} \right)^{-w} &  (w<0) 
\end{eqnarray}
In fact, if we ignore magnetostatic interactions, $\tilde F_w$ and $i \tilde F_w$ are the exact lowest-frequency normal modes with
winding number $w$, differing only in phase. [There are also modes with radial and z-direction
nodes, but these have much higher frequency.]  These functions are sketched for $w=0$ (uniform mode) and $w=\pm 1$ (vortex and antivortex modes)
in Fig. \ref{figure:Modes}.

In the presence of magnetostatic interactions, the
radial and $z$ dependences change, but the azimuthal symmetry remains the same, so we can
measure the amplitudes of these modes by the (complex) moments
\begin{eqnarray}\label{mw}
\tilde m_w & \equiv \int {(x - iy)^w (M_x  + iM_y )dxdydz} \nonumber \\
 & = \int {\tilde F_w^* (x,y,z)(M_x  + iM_y )dxdydz}
\end{eqnarray}

\begin{figure}[!htb]
\begin{center}
\includegraphics[width=3 in]{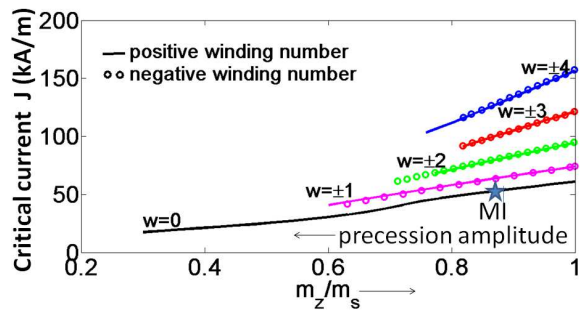}
\end{center}
\vspace{-0.2 in}
\caption{\label{figure:Jc} "Critical current" $J$ of normal modes\cite{munira}, continued to finite amplitude, labeled by winding number (circles are positive $w$, line is negative $w$, but these seem to be nearly degenerate.)  "MI" indicates magnetostatic instability of the $w=0$ (uniform) mode.
In these simulations $M_s = 500$ kA/m, $\alpha = 0.1$ for rapid convergence, $H_K = 1000$ kA/m, exchange $A = 10^{-11}$ j/m, $R = 30$ nm, $t = 4$ nm, cell size = 4 nm. [From ref. \cite{munira}]}
\end{figure}

We have calculated the actual normal modes (in the presence of magnetostatic interactions the radial dependence is nontrivial) by starting the system with transverse magnetization given by Eq. \ref{Fw}.  The higher-frequency modes will damp out faster [as $\exp (-\alpha \Omega_w t)$, where $\Omega_w$ is the frequency of the mode with winding number $w$, proportional to the critical current given in Fig. \ref{figure:Jc}], and we can suppress lower-frequency modes by projecting them out explicitly.  We keep the amplitude of the desired mode from decaying or growing by adjusting the spin torque. By definition, the normal modes are infinitesimal perturbations on the equilibrium "flower" state, but we can calculate periodic motions by this procedure for arbitrary amplitude -- the resulting critical spin torques (which are proportional to frequency) for the lowest few modes are shown in Fig. \ref{figure:Jc}.

\begin{figure}[t]
\begin{center}
\includegraphics[width=3in, height=1.0 in]{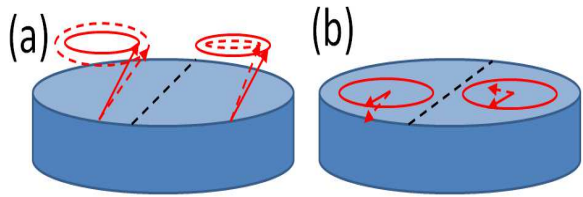}
\vspace{-0.1 in}
\caption{\label{figure:MI} Cartoon of (a) quasi-uniform state (solid arrows and precession circle) and perturbed by largest-eigenvalue eigenvector (dashed arrows and precession circle), for small precession angle; (b) the same for $90^\circ $ precession angle, where instability is easier to understand. [From ref.\cite{munira}}
\end{center}
\vspace{-3mm}
\end{figure}

A problem arises in calculating the large-amplitude quasi-uniform ($w=0$) orbit -- there is a magnetostatic instability, shown schematically in Fig. \ref{figure:MI}(a).
The reason for the instability is easiest to see when the precession is in-plane (Fig. \ref{figure:MI}(b)), when the perturbation tilts the magnetization upward at the right and downward at the left.  This clearly lowers the anisotropy energy \textbf{and} the magnetostatic energy, analogously to stripe domains in an extended film\cite{fujiwara}, so clearly is unstable if exchange is weak.

We can relate this instability, in which the left side is perturbed oppositely to the right side, to the winding-number modes in Fig. \ref{figure:Modes} -- if you add the vortex and antivortex modes, they cancel along the center line and point in opposite directions at the left and right. Thus the instability is essentially $\tilde F_1 + \tilde F_{-1}$.  This is a symmetry-breaking instability, which we can avoid numerically by projecting onto the correct symmetry, but in a real system it will grow exponentially.  We performed a Lyapunov analysis of this instability\cite{munira}, and found that one of the Lyapunov eigenvalues passes $1$ (indicating instability) at an angle of about $30^\circ$, corresponding to $U \approx 0.25$, for the parameters used in Fig. \ref{figure:Jc}.

\begin{figure}[!htb]
\begin{center}
\includegraphics[width=3 in]{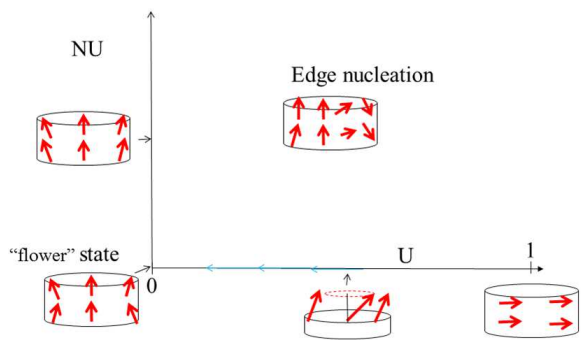}
\end{center}
\vspace{-0.2 in}
\caption{\label{figure:U-NU} Cartoon of the U-NU phase plane.  The minimum-energy flower state lies at the origin, and quasi-uniform precessional motions are on the horizontal axis. A general non-uniform state will lie in the interior of the graph; one with edge nucleation is depicted.
Horizontal axis indicates precession amplitude as in Fig. \ref{figure:Jc}, but direction is reversed.}
\end{figure}

Thus only three of the normal modes appear to be important at the instability -- one ($w=0$) corresponds to uniform precession, and is measured by the parameter $U$ (Eq. \ref{U}), $U = |\tilde m_0|^2 / m_s^2$.  It seems reasonable to measure the degree of non-uniformity near the instability by the sum of the squares of the two non-uniform modes:
\begin{equation}\label{NU}
NU = (|\tilde m_1|^2+|\tilde m_{-1}|^2)^2 / m_s^4
\end{equation}
(it turns out that we have to square it an extra time to make the equilibrium probability density (Fig. \ref{figure:Stoch}) smooth).
Thus we will plot the evolution of the system in the $U-NU$ phase plane shown in Fig. \ref{figure:U-NU}.

\section{Thermal Switching ($J<J_c$)}
\begin{figure}[!htb]
\begin{center}
\includegraphics[width=3 in, height=2 in]{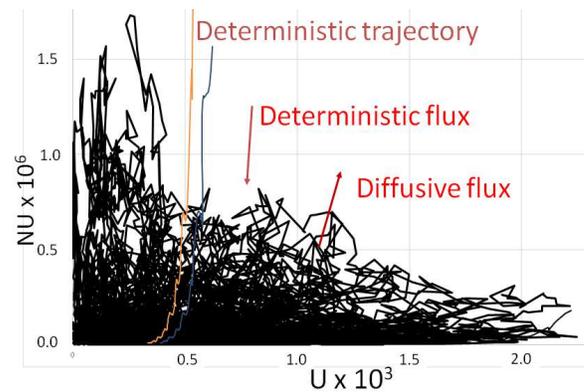}
\end{center}
\vspace{-0.2 in}
\caption{\label{figure:Stoch}
Stochastic switching trajectory at low current $J = 0.9 J_c$, in U-NU plane.  Arrows indicate approximate direction of deterministic and diffusive probability flux; in equilibrium these must cancel.  A few deterministic trajectories (with noise turned off) are shown (color online) -- the NU component relaxes faster than the U component, meaning that in the absence of noise the system returns quickly to uniform precession.
}
\end{figure}

To predict stability and read error rates, it is important to understand the switching rate for currents somewhat less than the critical current.  Fig. \ref{figure:Stoch} shows a stochastic trajectory in the $U-NU$ plane for $J=0.9 J_c$.  At this current there is no switching in the few tens of nanoseconds of this simulation -- it would be virtually impossible to get good statistics for the switching rate by brute-force simulation.  The system has a steady-state distribution which is close to Gaussian in the mode amplitudes.

The best way to determine probabilities of rare events is to construct and solve a Fokker-Planck (FP) equation to evolve the probability distribution in time -- this has been done frequently for 1D problems\cite{rate} and it is what we did\cite{A&V},\cite{V} for the single-macrospin spin-torque system (a 2D FP equation).  However, a many-macrospin model has thousands of degrees of freedom, so a complete FP treatment is not presently possible, and we cannot yet quantitatively calculate switching rates.  In the present paper we would like to do something less ambitious, namely, to determine whether edge-nucleation or precession instability is the dominant switching mechanism.

Instead of following a large ensemble of systems via a Fokker-Planck equation to determine precise probabilities of various trajectories, we will follow trajectories in a small ensemble by direct simulation.  Any attempt to deduce quantitative probabilities from our results would have huge statistical uncertainties, but if there is a qualitative conclusion to be found (i.e., if one of the two mechanisms was orders of magnitude more likely than the other), this should be evident from analysis of a small ensemble.  Our ensemble is not random, but is biased by our objective of finding edge-nucleation trajectories if they exist.  We would therefore expect edge-nucleation to be over-represented -- the fact that no trajectories that switched by edge-nucleation were found suggests strongly that they are very rare in an actual unbiased ensemble.

\begin{figure}[!htb]
\begin{center}
\includegraphics[width=3 in, height=2 in]{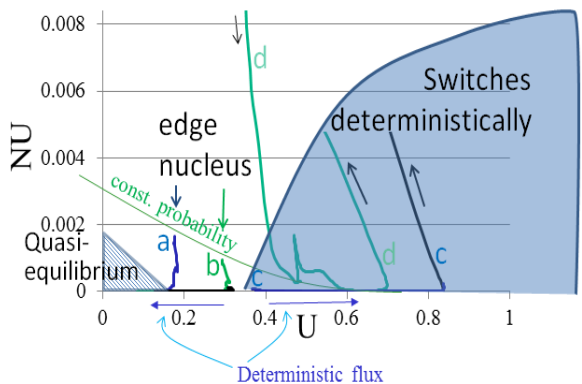}
\end{center}
\vspace{-0.2 in}
\caption{\label{figure:Trajs}
A small ensemble of deterministic trajectories, for $J<J_c$.
}
\end{figure}

Fig. \ref{figure:Trajs} shows such a collection of trajectories, for $J<J_c$ so that a deterministic trajectory that starts near the origin will relax toward the origin -- switching occurs only due to fluctuations.
However, there is a critical amplitude $U_{crit} \approx 0.35$ beyond which the system will switch - the deterministic flux points to the right, as indicated in the figure.  Note that the scale is very different from Fig. \ref{figure:Stoch} -- the steady state distribution would be squeezed into the shaded region near the origin labeled "Quasi-equilibrium".  But rare fluctuations will bring it out of this region, and we have tried to set up initial conditions that might lead to edge nucleation.  We can create a tilt at the edge by adding the vortex and antivortex mode to the uniform mode -- the trajectory labeled "d" started at $\tilde F_0 + 0.5 \tilde F_1 + 0.5 \tilde F_{-1}$, and is almost in-plane at the edge, and "a" and "b" started with similar but less extreme edge tilts.  The effects of noise on these trajectories are negligible, so we have omitted noise for simplicity. The edge nucleus does not grow in any of these trajectories -- all fall quickly back to uniform precession (the $U$ axis).  The uniform tilt $U$ of "a" and "b" was less than the critical value, so after returning to uniform precession they fall back to the origin.  The tilt of trajectories "c" and "d" was enough to cause them to switch, but not by edge nucleation -- they return to uniform precession, which enlarges until they hit the magnetostatic instability and $NU$ begins to grow exponentially, at the right side of the graph.  This creates a domain wall at the center, whose motion completes the switching deterministically.

\section{Overdriven Case ($J>J_c$)} \label{section:overd}
\begin{figure}[!tbh]
\begin{center}
\includegraphics[width=3 in, height=2 in]{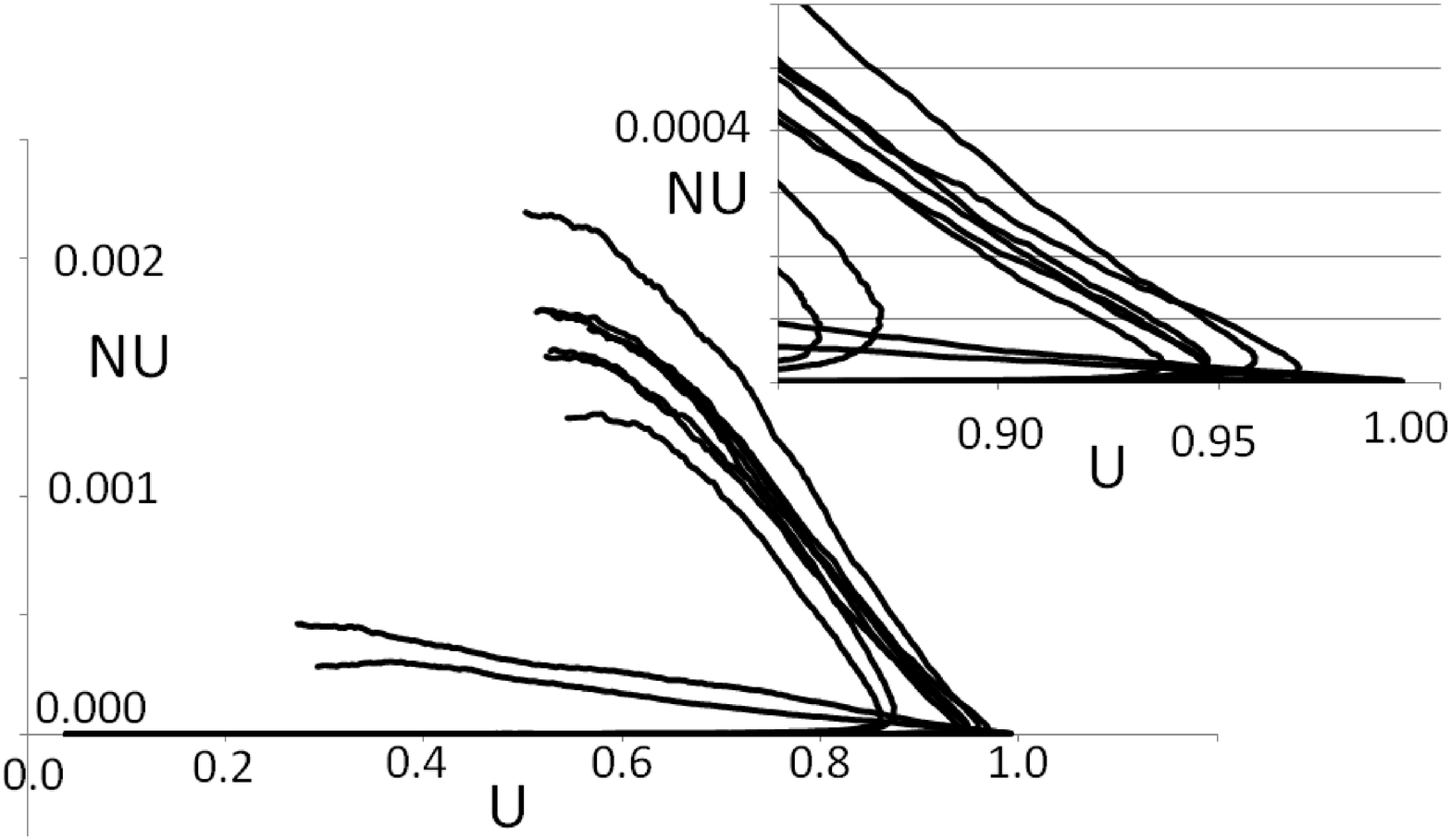}
\end{center}
\vspace{-0.2 in}
\caption{\label{over}
Stochastic switching trajectories at high current $J = 2.5 J_c$.   All trajectories begin at the lower left with the same initial condition. and move to the right (coherent
precession amplitude U increases).
To avoid too much crossing, each trajectory is truncated when NU starts to decrease.
Inset: Detail of region near the equator ($U=1$).
}
\end{figure}

In Fig. \ref{over}, we show several stochastic trajectories with high current.
The random non-uniform amplitude happens to be small for two of them, so they remain nearly coherent as they pass the equator ($m_z=0, U=1.0$) and reverse (moving to the left with small nonzero $NU$ in this plot, but in the southern hemisphere).  In the others, the non-uniform amplitude is larger, so its exponential growth makes them incoherent before reaching the equator, and the final stage of switching resembles domain wall motion (not shown).

\section{Conclusion}
We have developed a way of visualizing non-uniform switching, which makes it possible to distinguish between different switching modes, such as edge nucleation and the magnetostatic instability we have previously described.  We find that for thermal switching, with currents near the critical current $J_c$ for spin-torque switching, despite our best efforts to find an initial condition that leads to edge nucleation, switching almost always occurs via the magnetostatic instability.

\end{document}